\documentclass[modern]{aastex631}
\usepackage{hyperref}

\begin{document}

\title{Educated towards research: the first five years of the undergraduate mentoring program at Konkoly Observatory\footnote{This work has been published first in the journal \href{https://mersz.hu/hivatkozas/matud202301_f84002}{Magyar Tudomány} in Hungarian \citep{orig}. This is the slightly edited English translation of the original, produced with permission. }}

\correspondingauthor{L. Moln\'ar}
\email{molnar.laszlo@csfk.org}

\author{L\'aszl\'o Moln\'ar}
\affiliation{Konkoly Observatory, Research Centre for Astronomy and Earth Sciences, E\"otv\"os Lor\'and Research Network (ELKH), Konkoly Thege Mikl\'os \'ut 15-17, H-1121 Budapest, Hungary}

\author{L\'aszl\'o L. Kiss}
\affiliation{Konkoly Observatory, Research Centre for Astronomy and Earth Sciences, E\"otv\"os Lor\'and Research Network (ELKH), Konkoly Thege Mikl\'os \'ut 15-17, H-1121 Budapest, Hungary}

\author{R\'obert Szab\'o}
\affiliation{Konkoly Observatory, Research Centre for Astronomy and Earth Sciences, E\"otv\"os Lor\'and Research Network (ELKH), Konkoly Thege Mikl\'os \'ut 15-17, H-1121 Budapest, Hungary}



\begin{abstract}

In 2017 the Konkoly Observatory in Budapest published its first call for application inviting university students to carry out financially supported supervised research work and observing duties. The initiative quickly became popular and, so far, the program has supported 37 students. Five years later, is now time to summarize the experience gathered from both the institute and the participants. Notable results include numerous OTDK (student project) prizes awarded, first papers published, and acceptances into MSc and PhD programs both domestically and abroad, thus laying the foundations for the careers of several students. Among the feedback we have received from the students is the need for a more complex mentoring program, over and above of the funded research opportunities. A survey we conducted among the students indicates that communal and educational events are in the greatest demand, probably also induced by the lockdown restrictions experienced in the last few years. Through such events the students would not only build their community and start professional collaborations, but also learn more about various aspects of academia. 
In light of these results, we review possible avenues to improve the program.

\end{abstract}



\section*{The origins and beginnings of the program} 

The reasons and objectives behind the first call of the program for undergraduate students were quite simple. Konkoly Observatory of the Research Centre for Astronomy of Earth Sciences (CSFK), located in Budapest, Hungary, operates a number of telescopes and instruments located at its Piszkéstet\H{o} Mountain Station, in the M\'atra Mountains, about 100 km away. Although the telescopes underwent major modernization in the 2010s and can now be remotely controlled, there is still a need for a service astronomer on site who might even run an observing program themselves. Therefore, the recruitment of students started primarily in the context of training of the next generation of service astronomers at Piszkéstet\H{o}. Accordingly, the first published call included the learning of telescope operation through regular trips to the M\'atra mountains as a compulsory element, in addition to participating in research work. While these were the practical reasons for recruitment, the inspiration for organising it as a call for ``demonstrators''\footnote{The position is called  demonstrator, in Hungarian, a term originally used for students and teaching assistants who participate in experimental demonstrations at universities.} came from the personal experiences of L\'aszl\'o Kiss, the then Director of Konkoly Observatory, from his time as a physics demonstrator at the J\'ozsef Attila University of Sciences (now the University of Szeged). The program has continued to enjoy the unwavering support of the management of the CSFK research centre and has more recently been under the supervision of the current Director of the Observatory, R\'obert Szab\'o. 

\begin{figure}
\includegraphics[width=1.0\textwidth]{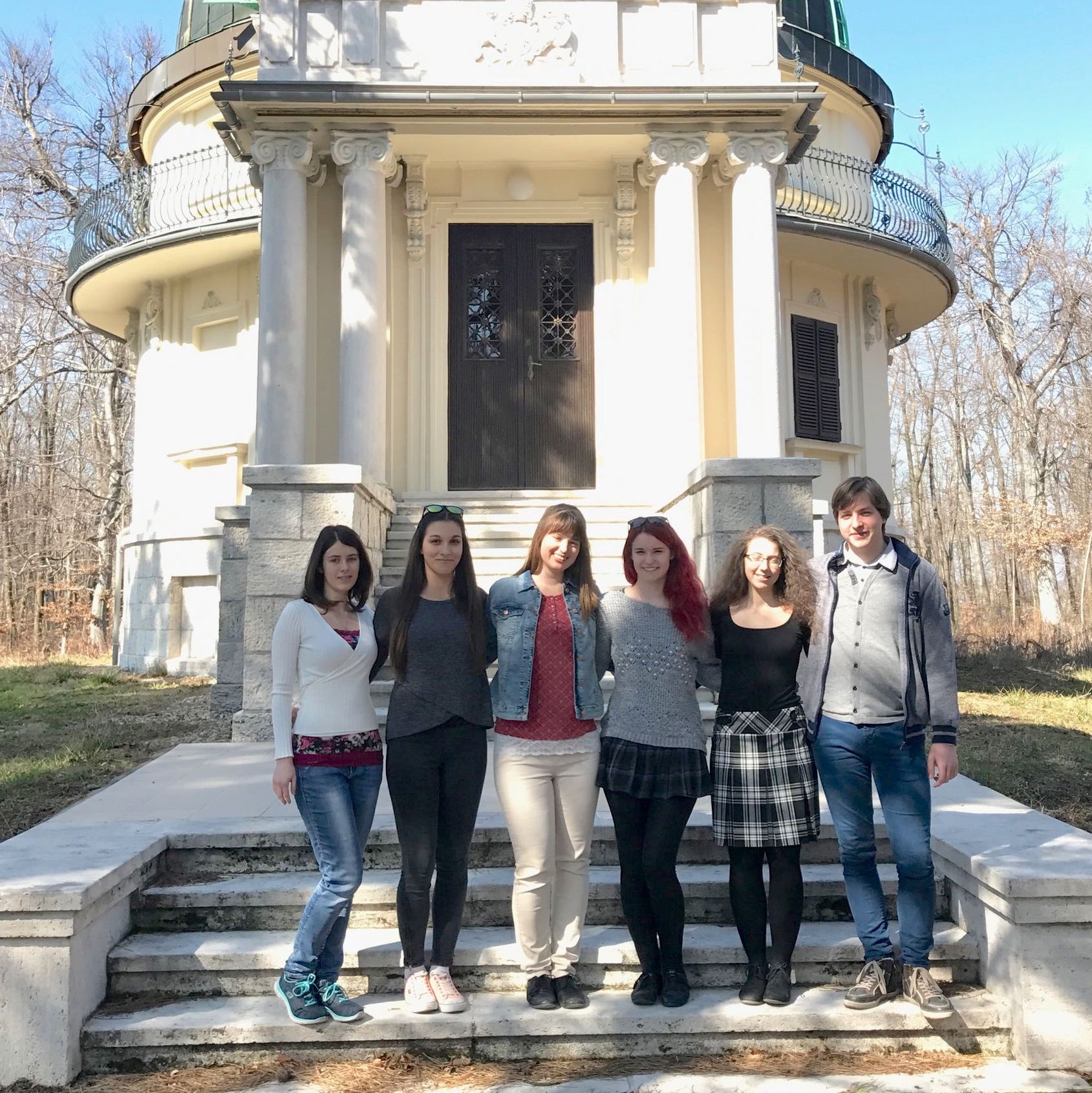}
\caption{How it all began: the first cohort of undergraduate students in 2017. (Photo: L\'aszl\'o Kiss) 
\label{fig:1}}
\end{figure}

The first cohort of students enrolled in the program was formed at the beginning of the spring semester of 2017, with one male and five female students (Figure~\ref{fig:1}). Since then, the program expanded rapidly: starting in the fall of 2017 the Observatory announced applications for 5 or 11 months, synchronized with the academic year, and funded either from grants or from the central budget, depending on the topic. In addition to the initial call, which focused on observational astronomy, two other calls were subsequently launched, reflecting the diversity of the research and development projects at Konkoly: one for numerical astrophysics (since 2018) and one for engineering and instrument development (since 2019). The research topics from which students could chose were proposed by Konkoly researchers. The number of applicants selected for each call rose from the initial six to ten, and then to fourteen, with many of the students continuing in the program for more than 11 months. The call proved to be very popular, regularly oversubscribed by one and a half to two times.

\begin{figure}
\includegraphics[width=1.0\textwidth]{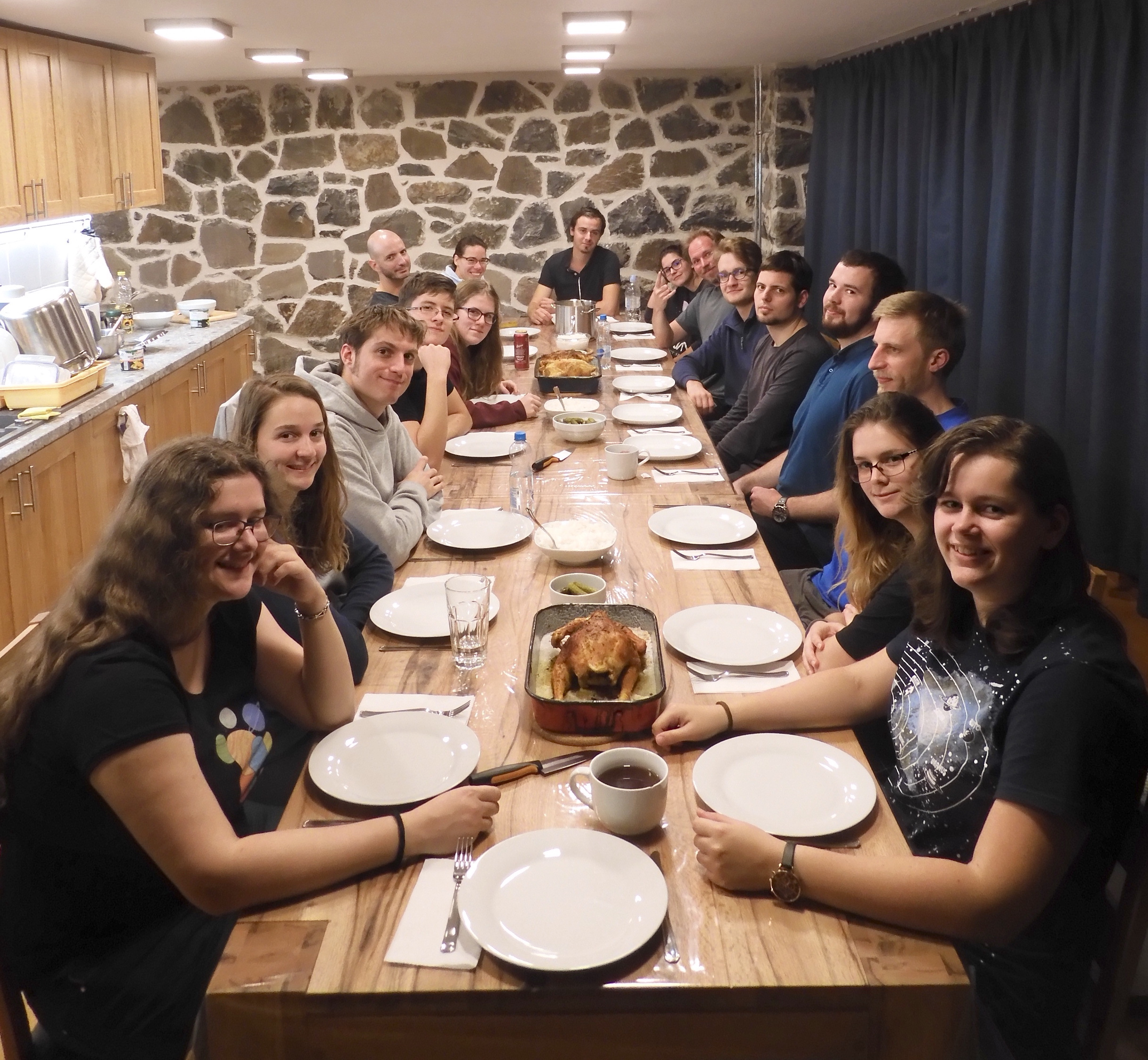}
\caption{How it is going: the cohort of 2021, plus some postdocs and staff scientists during the autumn retreat at the Piszkéstet\H{o} Mountain Station. (Photo: Csilla Kalup) 
\label{fig:2}}
\end{figure}

In the first five years of the program, i.e., between 2017 and 2022, a total of 37 students were offered the opportunity to participate and be financially supported by the program\footnote{Since the writing of the original article, selections have been made for the 2022/2023 year, funding 14 students, six of whom are new and not included in the statistics reported in this section.}: 29 in observational astronomy, four in numerical astrophysics and four in engineering (Figure \ref{fig:2}). The gender distribution of the admitted students is about 60--40\% male to female, although, if we consider the number of semesters sponsored, this ratio changes to 53--47\%. Overall, this is in good agreement with other statistics that show that the gender ratio in science undergraduate education is close to parity and that only at later career stages, typically around and after PhD, does the proportion of women researchers start to decline (see e.g., \citealt{doi/10.2777/06090}).

\section*{Is it worth employing students for research?}

It is worth asking what the return on investment for a non-educational institution might be in employing undergraduate or postgraduate students to do research. For example, would it not provide a better return on investment and more reliable outcomes to fund fewer young researchers or post-doctoral positions with the same budget instead? After all, the main task of students is to complete their education, and universities also provide them with study and research stipends (albeit sometimes rather token amounts). Furthermore, academic institutions have been already been involved in the scientific training of students in the past, even in an organized way, in Hungary through the TDK and OTDK (University and National Scientific Student Associations' Conferences\footnote{The TDK and OTDK conferences are annual and biennial events at the faculty and national levels, respectively, where students can present their research in a competitive environment, and receive awards for them.}, although without providing any financial support. 
 
In the context of to the issue discussed in this section, the fact that students have not received financial support in the past may not be a strong reason for continuing to do so. It is simply an appeal to tradition\footnote{\textit{Argumentum ad antiquitatem}: a type of logical fallacy that assumes the correctness of its premise by stating something along the lines of ``it's always been done this way".}, implicitly assuming that the old ways are not worth changing. However, if a research institute has the means to support students financially, why not reward the research they carry out in addition to their undergraduate studies? We are increasingly confronted with students taking part-time jobs in labor-intensive occupations (e.g. fast-food outlets and other student jobs) to support themselves. If these jobs and incomes could be replaced with paid participation in academic research, everyone involved would benefit. We can look at the problem from the other side: students spend a lot of time and additional resources on extracurricular activities outside their studies, which could otherwise be spent on other income-generating activities. By providing paid internships for demonstrators, we could instead allow them to focus on science with more financial security. While not all students will continue their careers at the same institution or even in the same profession, such financial support can be a critical factor in keeping applicants on their career paths in their professions. In this regard, the demonstrator program also helps to reduce the extent to which differences in students' financial resources affect their ability to engage in research.  
 
In addition to financial support, professional development also pays dividends for the institution by training more experienced young researchers. Research is a complex activity, with many interdependent steps, and is not equivalent to laboratory exercises, where predefined tasks must be performed correctly to achieve a known outcome. In the case of a research task, the hypothesis generation, testing, evaluation of the results and their discussion with the supervisor are themselves all part of the task. Students who already have this knowledge will have an advantage when writing proposals, applying for doctoral programs (even internationally) and will be more likely to pursue a successful research career.  
 
How can we measure the impact of the program so far on our latest generation of astronomers? Although we have seen many positive results in recent years, it is actually very difficult to objectively measure how many students and to what extent has the program helped to start their research careers. As there are many other influences on individual careers,  immediate results are not strong predictors of future careers in science. Even at the international level, few studies have attempted to assess the benefits of undergraduate research compared to, for example, with traditional laboratory practice \citep{linn-2015}. Nevertheless, it is still worthwhile to summarize below the successes so far.

\section*{From TDK participation to the pages of Science}

As the Konkoly demonstrator program has been running for five years now, we have had the opportunity to review the results achieved so far by its participants. In Hungarian higher education, one of the main venues for The presentation of research work carried out alongside studies is the aforementioned institutional and national Scientific Student Associations’ Conferences (TDK and OTDK) mentioned above. Since the start of the demonstrator program, two OTDKs have taken place, with three and four sessions on astronomy in 2019 and in 2021, respectively. While it is not yet meaningful to look for trends based on two data points, but it is worth noting that while in 2019, our students won one first, one second, and one third place plus four special prizes, in 2021, they won three firsts, three seconds, and three special prizes.
 
In addition, there is much to be gained from being actively involved in research beyond a good TDK or OTDK result. The most straightforward form of this is for a student to participate in data collection, for example, and then co-author a paper reporting the results. This is not always self-evident: research assistants and lab assistants do not necessarily become co-authors in all disciplines, and this has not always been the case in astronomy, either, although it is becoming more common. Indeed, for astronomers, the data collected from the sky can be essential, and it is common practice to reward the work of an observer working at the telescope with authorship, even if they are only students in training.
 
To examine authorship, we have collected the articles written since the beginning of the program that list our students as authors\footnote{\url{https://ui.adsabs.harvard.edu/public-libraries/lw4yhg3dQai-C5Epcxn8Gg}}. Of course, it is difficult to determine in retrospect which articles were directly related to the demonstrator tasks. We therefore drew a simple, though probably imprecise, line and considered articles published since the demonstrator application and including the two years after its completion. We identified a total of 77 articles, 63 of which were published in peer-reviewed journals, with the remaining 12 being conference proceedings and other smaller publications. These received a total of 708 citations by the time of writing this manuscript (and up to the  translation of the original in January 2023). This equates to almost two papers and 16 citations per student, although the distribution is far from even. The data also show that four of the six most cited papers use observations collected with telescopes at Piszkéstet\H{o} Mountain Station of Konkoly Observatory, which has been classified by the National Research, Development and Innovation Office as one of the TOP50 research infrastructures in Hungary \citep{nkfih}. This clearly shows that the original goal of the program has been successfully achieved.
 
It is also important to note that student participation in publications is not always limited to small contributions. We identified eighteen first-author papers from ten students, some of which, of course, were already part of their subsequent doctoral research. However, in at least eleven cases we were able to confirm that the students had written their journal article during their undergraduate studies, under the guidance of their respective supervisors \citep{bora-2022,csornyei-2019,kalup-2021-2,kalup-2021,kecskemethy-2023,konyves-toth-2020,krezinger-2020,krezinger-2020-2,seli-2019,szabo-2021,veres-2021}.
 
Many of our participants are currently pursuing doctoral studies, either in Hungary or at prestigious institutions abroad. In addition, several have successfully enrolled in Master's programs abroad. Of the original six participants, R\'eka K\"onyves-T\'oth and Gabriella Zsidi have already defended their doctoral theses. But perhaps the most astonishing achievement of the last five years is that two of our Bachelor students, Blanka Vil\'agos and Benj\'amin So\'os, as members of the RADIOSTAR ERC research group\footnote{\url{https://konkoly.hu/radiostar/}} on nuclear astrophysics, became co-authors of a paper published in Science \citep{cote-2021}. This example illustrates the benefits of managing a research team in an inclusive way, by bringing together junior and senior members in joint projects according to their expertise.  

\section*{Halfway from student scholarships to a mentoring program}

Supporting undergraduate students also provides an opportunity to bring together young people involved in the work of the Observatory beyond their individual project with a supervisor. We have already taken some initiatives in this direction, such as a three-day kick-off workshop in early 2020. There, undergraduate and doctoral students could work together in a focused way on the tasks of their choice and present their progress to each other over the three days. However, like so much else, these programs were made impossible by the COVID-19 pandemic, and for a time the demonstrator program was reduced to a series of online video-consultations. 
 
The cohort of 2021/2022 was the first to have the opportunity to hold community programs again. In October 2021, the whole group spent several days in Piszkéstet\H{o}, learning about telescopes and observing (Fig.~\ref{fig:2}). This also brought up the question of what the objective of the demonstrator program is and whether it has changed since five years ago. Are we still training observers for telescopes? Or are we providing grants for research work worthy of a thesis, an OTDK work, and/or a paper? Has Konkoly Observatory laid the groundwork for a more comprehensive mentoring program?

We are confident that the latter will be true. Seven years ago, an article was published in Magyar Tudom\'any (the journal where this paper originally appeared) summarizing the vision of Hungarian astronomy for the second half of the 2010s \citep{csub-2016}. The article dealt only relatively briefly with the issues of recruitment and training of researchers. However, there has been an increasing emphasis on more support for human resources, be it mentoring, support for young researchers or gender equality by the 2020s, compared to the mainly infrastructural development goals listed in that paper. There is a growing global need for institutions to support young researchers, from students to postdocs, not only with gestures and opportunities, but also with more tangible forms of support, from mentoring to financial aid and even mental health care. The development of a more complex mentoring program that supports the careers of prospective and young researchers in multiple ways, fits perfectly into this philosophy.
 
In order to get a more accurate picture of how the supported students felt about the demonstrator program and what new elements they would like to see as part of a more comprehensive program, one of us (LM) conducted a short survey among them. We received 11 responses to the questions: all the students, without exception, were very satisfied with the demonstrator job and considered it very important for their professional development. They were also satisfied or very satisfied with their supervisors, and most students indicated that they would apply for the program in the next academic year. Responses were mostly scattered on the importance in terms of livelihood, with most responses ranging from moderately important to highly important. This is consistent with the different financial backgrounds of the individuals and shows that there is indeed an equalizing, equity-enhancing impact of the program.
 
In addition to conducting the research projects themselves, most students were also advised by their supervisors to write a TDK thesis, to become an author of a journal article, or to collaborate with a research group or other researchers. Attending a conference or summer school in person was the least common option, for a variety of reasons, ranging from the constraints of the COVID-19 pandemic to the financial burden of travel. However, hybrid and online events are now often available at low cost, making it worthwhile to increase the promotion of these opportunities to students.
 
The importance of networking and community building is also reflected in the fact that community building was the most valued aspect of the week spent at Piszkéstet\H{o} in 2021, with higher scores than learning about instrumentation and astronomical observations. This is understandable in a perspective where universities, forced to provide distance learning due to the pandemic, have almost completely lost their community-building roles, so the value of alternative options has increased significantly.
 
\begin{figure}
\includegraphics[width=1.0\textwidth]{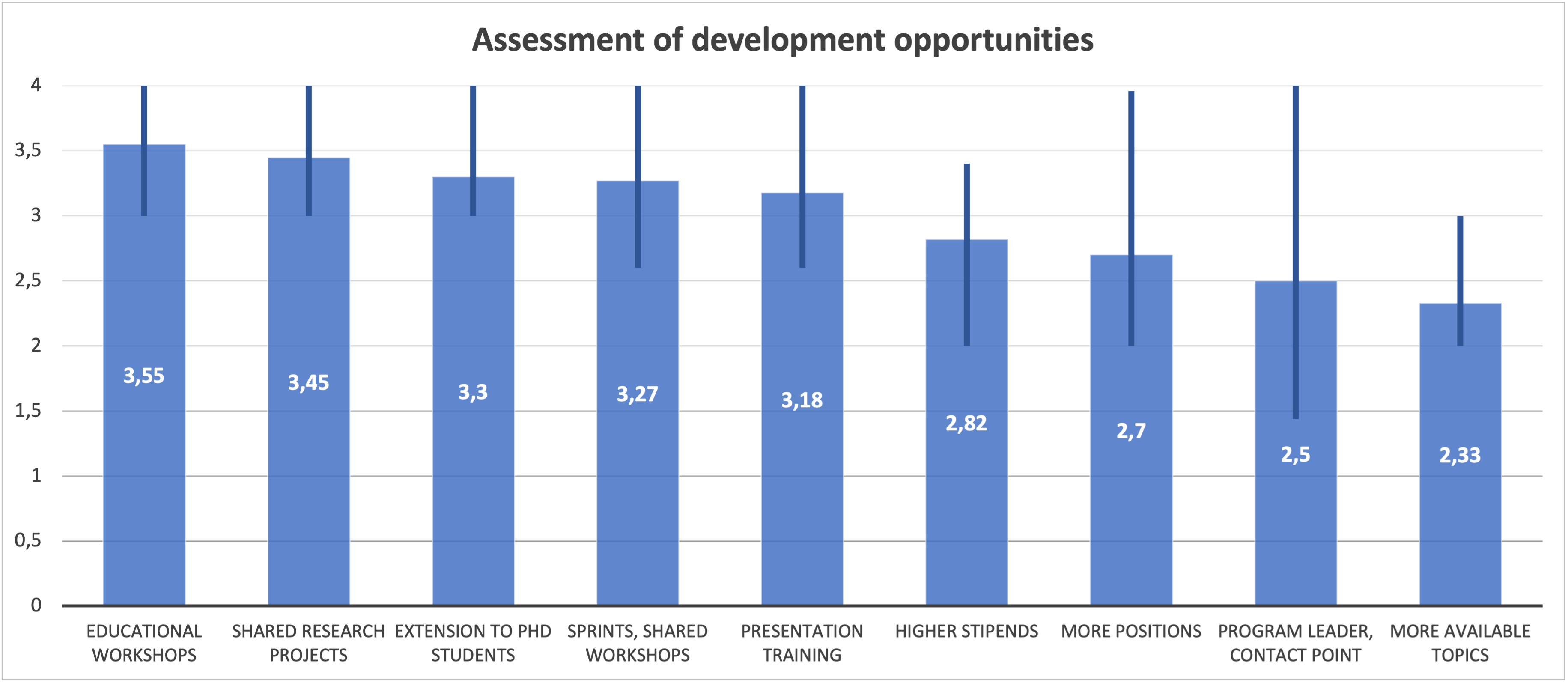}
\caption{Students' evaluation of various possible development opportunities suggested for the program. The questionnaire was filled out by 11 students, who were asked to rate the importance of each opportunity on a scale of one (not at all important) to four (highly important). 
\label{fig:3}}
\end{figure}

Perhaps the most important results came from the questions related to the development of the program (Figure \ref{fig:3}). Among the future possibilities, the lowest score was given to \textit{availability of more topics}: in fact, there were already a large number of research topics to choose from. Interestingly, however, the seemingly trivial option of \textit{higher stipends} did not receive the highest score: the \textit{shared research projects} and \textit{extension to PhD students} options both scored better. The most voted option was the possibility of \textit{educational workshops}, where education is understood here mainly as knowledge transfer related to research and career development. The feedback suggests that students find it difficult to access detailed information on topics such as doctoral training and subsequent career paths; preparing professional publications; opportunities and conditions for attending conferences; types of research grants, and how to obtain and implement them.

\section*{Possible ways forward}

In light of the above, there is clearly room for a more comprehensive mentoring program, in addition to the current professional activities and financial support. Community programs and educational sessions were identified by the students as a top priority and should be considered as soon as the pandemic situation allows. It is also worth considering whether research topics can be linked together to form collaborative projects. In addition to training, more personalized mentoring focusing on application and proposal writing could be added to the program, with experienced researchers helping to select and write appropriate proposals.

Furthermore, mentoring and helping students does not have to end when they graduate. This is especially true for those who decide to pursue a doctorate in Hungary. At present, the stipends offered by the doctoral schools that train astronomers\linebreak (140 and 180 thousand HUF/month for the first and second two years respectively; equivalent to annual stipends of 4200 and 5400 USD) are clearly not enough to cover the living costs of a young researcher in their mid-twenties who wants to start an independent life. A doctoral student should already be concentrating full-time on training and research, but with a second job to make ends meet, this can become much more difficult and may even lead to career drop-out.

Until the situation of Hungarian doctoral stipends is resolved, there are relatively few options available: joining a research grant awarded to someone else at the right time; short-lived proposals that pop up from time to time, such as the Cooperative Doctoral Program (KDP), which has so far been announced only in 2020 and 2021; and the doctoral category of the New National Excellence Program (\'UNKP). In the case of the \'UNKP, the evaluation criteria focuses on OTDK rankings, publications, as well as on the doctoral research achievements and patents already completed. In other words, an early-stage doctoral student is only eligible if they have already carried out intensive research in addition to their university studies. These are further arguments in favor of both the mentoring program and its possible extension to doctoral students.

Of the written comments received in the survey, the following is perhaps the best summary of the experience so far:

\textit{``Without this program, I think far fewer people would stay in the field (make it to the doctorate). My suggestion is more community programs, which could result in a close-knit community of young people. I think that this could reduce the number of drop-outs later (in careers)."}

Therefore, rather than examining individual careers, it is probably more appropriate to think of the program in terms of its generational impact, looking back on it later, and reassessing the changes it has brought to Hungarian astronomy and the international recognition of Hungarian astronomers, even on a decade-long scale.


\vspace{0.5cm}
\paragraph{Acknowledgements} We thank Csilla Kalup and Attila Bódi for their efforts in organizing the Piszkéstet\H{o} and kick-off workshops and Maria Lugaro for proofreading this paper. We are grateful to all of the supervisors who gave their time and often their own funding to support the students.

\bibliographystyle{aasjournal}



\end{document}